\documentclass[fleqn,twoside]{article}
\usepackage{espcrc2}

\newcommand{\be}{\begin{equation}}
\newcommand{\ee}{\end{equation}}
\newcommand{\ba}{\begin{eqnarray}}
\newcommand{\ea}{\end{eqnarray}}
\def\ch{{\rm ch}}

\usepackage{graphicx}
\usepackage[figuresright]{rotating}

\newcommand{\AmS}{{\protect\the\textfont2
  A\kern-.1667em\lower.5ex\hbox{M}\kern-.125emS}}

\title{Two-dimensional gauge theories of the symmetric group $S_{n}$
and branched   
n-coverings of Riemann surfaces in the large-n limit.} 
\author{A. D'Adda\address[TO]{
INFN sezione di Torino, and Dipartimento di Fisica Teorica,
Universit\`a  di Torino,
I-10125 Torino, Italy} and
P. Provero\addressmark[TO]\address{Dipartimento di Scienze e
Tecnologie Avanzate, Universit\`a  del Piemonte Orientale, 
I-15100 Alessandria, Italy}}
       
\begin{document}

\begin{abstract}
\vspace{1pc}
Branched $n$-coverings of Riemann surfaces are described by a 
2d lattice gauge theory of the symmetric group $S_{n}$ 
defined on a cell discretization of the surface. We study the theory 
in the large-$n$ limit, and we find a rich phase diagram with first 
and second order transition lines. The various phases are 
characterized by different connectivity properties of the covering 
surface. We point out some interesting connections with the theory of 
random walks on group manifolds and with random graph theory.
\end{abstract}

% typeset front matter (including abstract)
\maketitle
\section{Introduction}
The problem of counting in how many ways a Riemann surface can be 
covered $n$-times without allowing folds but allowing branch points 
is equivalent to computing the partition function of a lattice gauge 
theory defined on a discretization of the surface, with the symmetric 
group $S_{n}$ as the gauge group. The choice of the action for the gauge theory determines the   
branch-point structure of the coverings one wants to 
count\cite{Kostov:1997bs,Kostov:1998bn,Billo:2001wi}. 
\par 
Therefore 2d $S_{n}$ gauge theory is of interest not 
only beacuse it is a simple but non-trivial non-abelian gauge theory, 
but also because it has a natural interpretation in terms of a 
2d string theory, that is a theory of maps from a 
2d world sheet to a 2d target space. 
Moreover the strings described by $S_{n}$ gauge theory, that can have 
branch points but not folds, are precisely the ones that enter the 
string-theoretic formulation of $SU(N)$ 2d gauge theory, discovered in 
\cite{Gross:1993hu,Gross:1993yt}.
Finally, branched coverings of Riemann surfaces appear naturally when 
one quantizes 2d Yang Mills theory in the unitary gauge. This 
gauge choice 
was shown in Refs.\cite{Billo:1998fb,Billo:1999ts} to 
actually define a different theory, with a natural 
intepretation in terms of 2d strings. 
\par
In this work we will consider 2d $S_{n}$ gauge theory on a genus-zero 
Riemann surface, and in 
particular the theory that corresponds to coverings with quadratic 
branch points only, in the 
large-$n$ limit, and study its phase diagram. It turns out that one 
can obtain a non-trivial phase diagrams by letting the number of 
plaquettes $p$ of the discretized target Riemann surface scale with 
the order $n$ of the coverings as
\be
p=A n \log n
\label{scaling}
\ee
For example if one considers the theory in which exactly one 
quadratic branch point is placed on each plaquette, one finds that 
at $A=\frac12$ there is a phase transition separating a phase in 
which the covering has more than one connected component from one in 
which the covering surface is connected. By allowing each plaquette 
to have either no branch point or one quadratic branch point with a 
given probability, the phase diagram becomes more complex, with lines 
of first and second order phase transitions.  
\par
These results can be obtained by writing the partition function of 
the theory, as it is customary in 2d lattice gauge theories, as a sum 
over irreducible representations of $S_{n}$, and then looking for the 
representations that give the highest contribution in the large-$n$ 
limit. This procedure is outlined in Sec. 2.
\par
Some insight into the nature of the various phase transitions can be 
gained from an exact correspondence, that we describe in Sec. 3, 
between 2d gauge theories on genus-zero Riemann surfaces and random 
walks on the gauge group. The correspondence is completely general 
with respect to the choice of the gauge group and the action. If one 
applies it to our gauge theory, one can map it into a particular 
random walk on $S_{n}$ that has been studied in the 
mathematical literature as an example of {\it cutoff 
phenomenon}\cite{Diaconis:1981}. In Sec. 
4 we will show that one of the phase transition lines we find 
corresponds to the cutoff phenomenon in this random walk.
\par
Finally, another precise connection can be made with the theory of random 
graphs, and precisely with what are called {\it phase transitions} in 
random graphs. In Sec. 5 we will describe this connection, and use it 
to argue  that the phases of our model are distinguished by the 
connectivity properties of the world-sheet. 
\par
Sec. 6 will be devoted to some final remarks. In this contribution we
just present our  
results: for more details and derivations, we refer the reader to 
Ref.\cite{D'Adda:2001nf}.
\section{The model and the phase diagram on the sphere}
The equivalence between 2d $S_{n}$ gauge theories on Riemann surfaces 
and the problem of counting branched coverings of such surfaces is 
described in detail in Ref.\cite{Billo:2001wi}. Such equivalence can 
be established for all possible branch point structure of the 
coverings. Here we will consider quadratic branch points only: On each 
plaquette of the target Riemann surface there is, with probability $x$, 
a quadratic branch point, or, with probability $1-x$, no branch point. 
\par
The partition function on a disk depends on the branch point type on the 
boundary, that is, in the language of the gauge theory, on the 
holonomy $Q\in S_{n}$ at the boundary of the disk, and can be written as
\ba
&&Z_{n,{\rm disk},p}(Q)\nonumber\\
&&\ =\frac{1}{n!} \sum_r d_r \ch_r(Q)
\left[ (1-x)+x\frac{\ch_r(\bf{2})}{d_r} \right]^p
\label{zetadisk}
\ea
where $p$ is the number of plaquettes, the sum runs over all the 
irreducible representations $r$ of $S_{n}$, $d_{r}$ is the dimension 
of the representation $r$ and $\ch_r(\bf{2})$ is the character of a 
transposition in $r$. The problem of computing the partition 
function  Eq.(\ref{zetadisk}) in 
the large-$n$ limit consists in finding which representations give the 
largest contribution to the sum when $n\to\infty$.
\par
The irreducible 
representations of $S_{n}$ are in one-to-one correspondence with the 
Young tableaux with $n$ boxes. The results of the analysis 
can be expressed in terms of the shape of the Young tableaux that 
dominate the sum in the large $n$-limit. We will restrict ourselves to 
the case of the sphere, where $Q$ is the identical permutation. 
\par
If one lets $p$ scale with $n$ as in Eq.(\ref{scaling}) one finds three 
phases in the $(x,A)$ plane, shown in Fig. 1:
\begin{itemize}
\item
In phase I, the Young tableaux that dominate the sum have both rows 
and columns scaling as $n^{1/2}$.
\item
In phase II the sum is dominated by Young tableaux made of a single row of 
length $2 A x$ attached to a part of area $(1-2 A x)n$ whose rows and 
columns scale like $n^{1/2}$.
\item
In phase III the sum is dominated by a Young tableau made of a single 
row (or column) of length $n$.
\end{itemize}
\begin{figure}[htb]
\vspace{9pt}
\includegraphics[height=5.5cm]{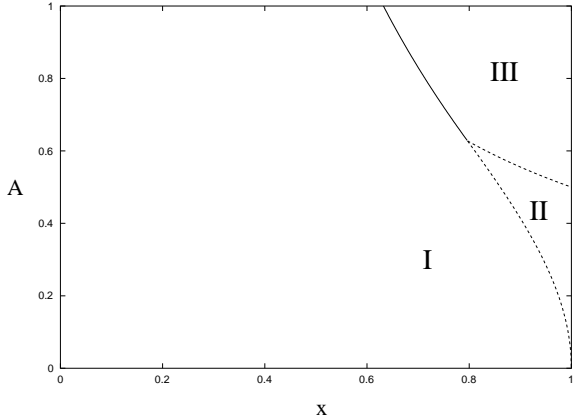}
\caption{The phase diagram of the model on a sphere.}
\end{figure}
The free energy in the large-$n$ limit can be written exactly, so that 
one can determine the order of the various transitions. It turns out 
that the (I,II) and (I,III) transitions are first order, while 
(II,III) is second order, characterized by a discontinuity of the 
second derivative of the free energy with respect to $A$. 
\par
What this type of analysis does not easily give is a 
characterization of the three phases in terms of some order parameter 
of clear physical meaning. This problem will be solved in part by 
mapping the theory first into a random walk on $S_{n}$, then into a 
classical problem of random graph theory. This will be done in the next 
two sections.
\section{2d gauge theories are random walks on the gauge group}
A two-dimensional gauge theory on a disk is equivalent to a random
walk on the gauge group manifold, the area of the disk being
identified with the number of steps and the gauge theory action with
the transition probability at each step.
\par
This result is completely general with respect to the choice of
the gauge group and the action, as is proven in Ref.\cite{D'Adda:2001nf}. 
Here we show how this result emerges in the $S_n$ gauge
theory defined by Eq.(\ref{zetadisk}) for $x=1$ (the extension to 
general $x$ is trivial), that is where all plaquette
variables are forced to be equal to a transposition.
\par
As shown by Eq.(\ref{zetadisk})
the partition function depends only on the holonomy $Q$ and the area
of the disk. 
It follows that we can freely choose any cell decomposition of the
disk made of $p$ plaquettes, {\em e.g.} the one shown in
Fig. 2.
\begin{figure}[htb]
\vspace{9pt}
\includegraphics[height=5.5cm]{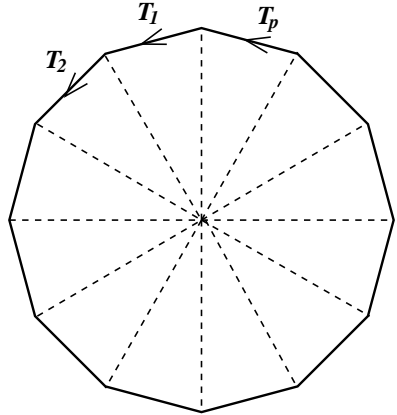}
\caption{A cell decomposition of the disk.}
\end{figure}
To compute the partition function is to count the ways in which we
can place permutations $P$ on all the links in such a way that
\begin{itemize}
\item
the ordered product of the links around each plaquette is a
transposition
\item
the ordered product of the links around the boundary of the disk is a
permutation in the same conjugacy class as $Q$
\end{itemize}
\par
Now we can use the gauge invariance of the theory to fix all the
radial links to contain the identical permutation. At this point the links
on the boundary are forced to contain transpositions: therefore
the partition function with holonomy $Q$ is the number
of ways in which one can write the permutation $Q$ as an ordered
product of $p$ transpositions. This in turn can be seen as a random walk
on $S_n$, in which, at each step, the permutation is multiplied by a
transposition chosen at random: the gauge theory partition function
for area $p$ and holonomy $Q$ is the probability that after $p$ steps
the walker is in $Q$.
\par
The {\it cutoff phenomenon} in random walks was discovered in
Ref.\cite{Diaconis:1981} by studying this random walk on $S_n$ (more
precisely the one corresponding to our model with $x=1/n$).  
It was shown in Ref.\cite{Diaconis:1981} is that if the number of
steps scales as $A n \log n$, then in the large-$n$ limit for $A>1/2$
the probability of finding the walker in any given element $Q\in S_n$
is just $1/n!$ for all $Q$: complete randomization has been achieved
and all memory of the initial position of the walker has been erased.
\par
In terms of the corresponding gauge theory, this statement translates 
into the following: for $A>1/2$, the partition
function does not depend on the holonomy $Q$, and at fixed holonomy 
$Q$, does not depend on $A$.  This is true in particular 
for  $Q=1$, corresponding to the
partition function on the sphere. Therefore the transition between 
phases II and III, found
in Sec. 2, has a natural interpretation as a cutoff phenomenon in the
corresponding random walk (see Ref.\cite{D'Adda:2001nf} for a
discussion about generalizing the result of Ref.\cite{Diaconis:1981}
to arbitrary values of $x$).
\section{A connection with the theory of random graphs}
The connection with the theory of random graphs is a consequence of 
the possibility,  established in the previous section, of mapping one 
of the phase transitions of our theory of coverings into a cutoff 
phenomenon of a random walk. In fact it was shown in Ref. \cite{Pak:2001} 
that some phase transitions in random graphs can be interpreted as 
cutoff phenomena in certain random walks that naturally generate 
ensembles of random graphs. 
\par
Consider the random walk corresponding to our model, taken now with 
{\it free} boundary conditions, and take for simplicity $x=1$:
at each step the permutation is multiplied by a random
transposition. From the point of view of coverings,
a step in which the transposition $(ij)$ is used corresponds to adding a
simple branch point that connects the two sheets
$i$ and $j$ of the covering surface. One can think of the process as 
the construction of a graph of $n$ sites, in which at each step one 
link, chosen at random, is added.  After $p=A n \log n$
steps the expected number of links is equal to the number of
steps (since the number of available links is $O(n^2)$ the fact that
the same link can be added more than once can be neglected in the
large $n$ limit; see Ref.~\cite{Pak:2001}).
\par
It is a classic result in the theory of random graphs
\cite{Erdos:1959,Erdos:1973}
that if the
number of links $p$ is smaller than $1/2\  n \log n$ then the graph is
almost certainly disconnected while for $p>1/2\  n \log n$ the graph
is almost certainly connected, where ``almost certainly'' means that
the probability is one in the limit $n\to \infty$. Therefore we 
conclude that the transition at $A=1/2$, for the model with free 
boundary conditions, is the point where the covering surface becomes 
connected. 
\par
Notice that the free boundary conditions are crucial for this argument
to work: in the case, say, of the sphere, the corresponding random
walk is forced to go back to the initial position in $p$ steps, so
that links in the graph are not added independently and the result of
Ref.\cite{Erdos:1959,Erdos:1973} do not apply. However there are 
strong arguments (see Ref.\cite{D'Adda:2001nf}) suggesting that   
connectedness of the covering 
surface can be used to distinguish the phases in the case of the 
sphere as well: phase III should correspond to a connected covering 
surface while phases I and II should not. 
\section{Conclusions}
We have shown that the two dimensional gauge theory of the symmetric 
group, that describes the statistics of branched coverings of a 
Riemann surface, has a remarkably rich phase structure in the 
large-$n$ limit. 
\par
The theory on the sphere can be studied by a 
variational approach that identifies which representations give the 
largest contribution in the large-$n$ limit. If the number of branch 
points is taken to scale as $n\log n$, three phases can be 
found, corresponding to different shapes of the dominant Young 
tableaux. The transitions between these lines can be located 
analytically and their order determined.
\par
All two-dimensional gauge theories on a genus-0 surface can be mapped
into random walks in the corresponding group manifold. In our case,
this allows us to interpret one of the transition lines as a cutoff
phenomenon in the corresponding random walk. 
\par
The theory on a disk, with free boundary conditions, can be studied
with methods of the theory of random graphs: this allows one to
show that there is a phase transition on a disk from a disconnected to 
a connected covering surface. From this one can argue, and with some limitations
prove, that the connectness of the covering is what characterizes the different
phases also on the sphere.
\par
This last result is particularly interesting since it provides a 
description of the various phases in terms of geometric properties of 
the covering surfaces, that is in terms of quantities that are 
natural when one considers the model as a two-dimensional string 
theory rather than as a Yang-Mills theory. Along this line, it was recently 
proposed in \cite{Semenoff:2001bi} that the phase transition we find is a 
close analogue of the Hagedorn phase transition in Matrix String 
Theory described in
Refs.
\cite{Grignani:1999sp,Grignani:2000zm,Semenoff:2000pu,Grignani:2001hb,Grignani:2001ik}. 

\end{document}